\documentclass[12pt]{article}
\usepackage{graphicx}
\usepackage{wrapfig}
\usepackage{epsfig}
\usepackage{epsf}
\begin{document}

\begin{center}
{\bfseries ABOUT ONE POSSIBILITY OF RELATIVISTIC DESCRIPTION OF
POLARIZED DEUTERON FRAGMENTATION } \\
\vskip 5mm \underline{L.S. Azhgirey}$^{1 \dag}$ and N.P. Yudin$^2$
\vskip 5mm {\small
(1) {\it Joint Institute for Nuclear Research, Dubna, Russia } \\
(2) {\it Moscow State University, Moscow, Russia } \\
$\dag$ {\it E-mail: azhgirey@jinr.ru
}}
\end{center}
\vskip 5mm

\begin{abstract}
In the framework of the light-front quantum theory developed by
Karmanov et al. an analysis of the experimental data on the tensor
analyzing power of the nuclear fragmentation of relativistic
deuterons with the large transversal momentum proton emission has
been made. With the Karmanov's wave function taken in system in
which z-axis directed along the deuteron beam we have managed to
explain  the existing data without invoking additional to nucleons
degrees of freedom. \end{abstract}
\vskip 8mm

The experiments with the polarized deuteron beams made in Saclay
\cite{arvieux}-\cite{punjabi} and Dubna \cite{ableev2}-\cite{azh3}
have led one to recognize that at the relativistic momenta of
deuteron there is something wrong either with the theory of
$A(d,p)X$ reactions or with the structure of the deuteron at short
distances between nucleons. At first, it was proved out that the
experimental dependence of the analyzing power $T_{20}$ on $k$ ---
internal momentum of nucleons in the deuteron --- does not change
sign at $k \sim 0.5$ GeV/$c$, as it followed from theoretical
calculations. Further \cite{azh3},  the pion-free deuteron-breakup
process $dp \rightarrow ppn$ in the kinematical region close to
that of backward elastic $dp$ scattering at a given value of $k$
depends on the incident momentum. This forces one to suggest that
in description of this quantity an additional variable is
required. This additional variable does not appears in the usual
schemes of calculations. At last, the recent measurements of the
tensor analyzing power $A_{yy}$ of the breakup of relativistic
deuterons on nuclei at large transverse momenta of emitted protons
\cite{large_t,azh5} show also that something unusual takes place
in the theory of this reaction since the measured $A_{yy}$-values
at fixed value of longitudinal proton momentum show a pronounced
dependence on the transverse proton momentum, that does not appear
in the calculations.

The theoretical considerations of $A(d,p)X$ reaction were carried
out in different lines and on the whole the situation with the
description of this reaction is contradictory. The most popular of
theoretical approaches is one of the light-front dynamics and in
this paper we follow it. In general this approach in the
approximation of a simple mechanism with the pole in $t$-channel
with using the standard deuteron wave functions satisfactorily
describes the differential cross section data \cite{ableev1,azh1}
(see, for example, \cite{azhrayu,azhigyu}). On the other hand, the
calculations of polarization observables in the same approach
\cite{azhyud1}, as a rule, does not reproduce the experimental
data; the exception is the paper \cite{kobushkin}, where the data
on the $T_{20}$ of nuclear fragmentation of relativistic deuterons
with the proton emission at 0$^\circ$ are described.

     The most simple statement would be to say that this
discrepancy between the theory and experiment is due to the
oversimplified mechanism of the reaction. But, in our opinion, at
the experimental accuracy achieved the possibilities of this
simple and thus valuable mechanism have yet not been exhausted.

\begin{wrapfigure}{r}{9cm}
\mbox{\epsfig{figure=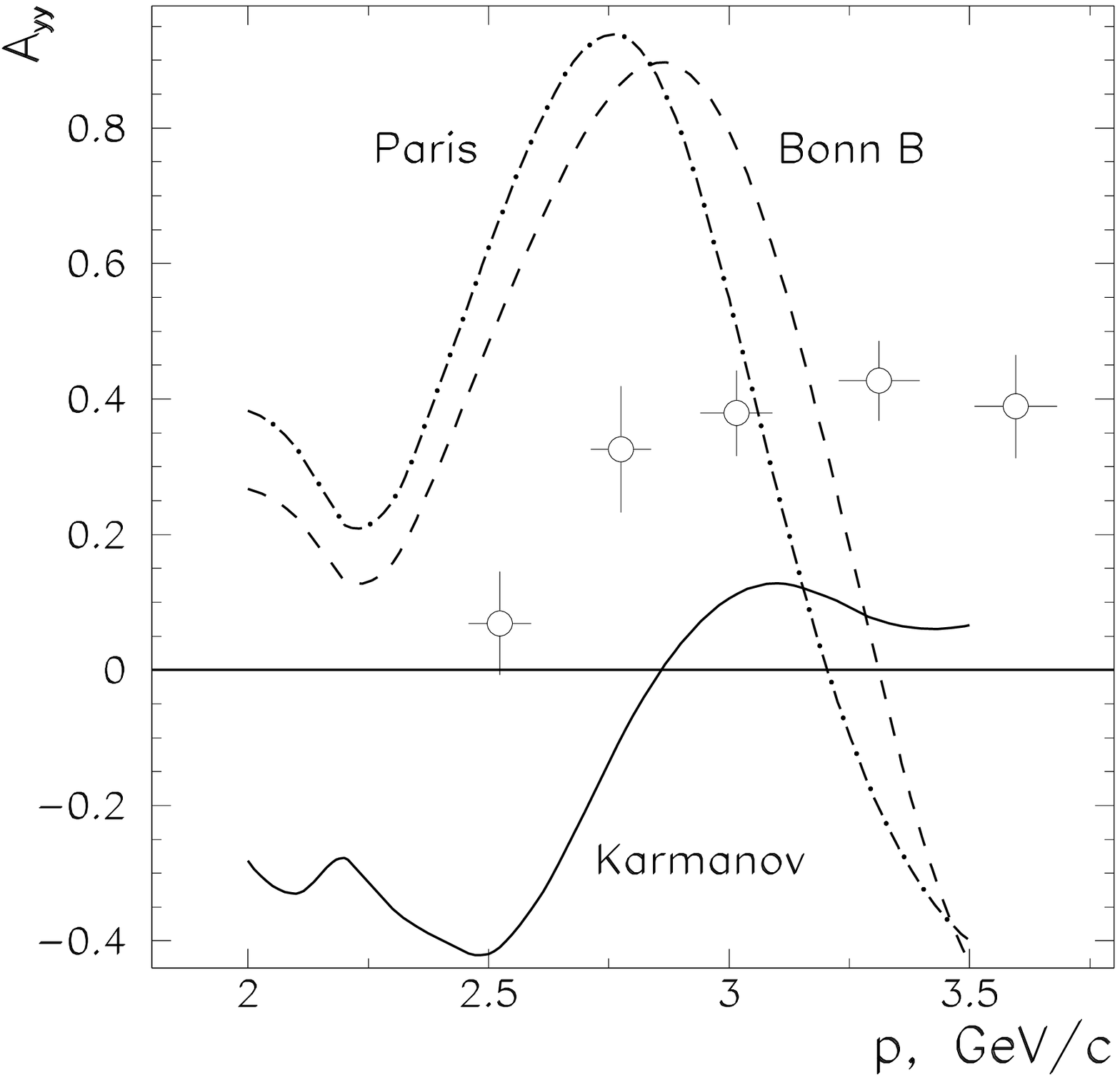,width=9cm,height=8cm}} 
{\small {\bf Fig. 1.} 
Parameter $A_{yy}$ of the reaction $^{9}Be(d,p)X$ at an initial
deuteron momentum of 4.5 GeV/$c$ and a proton emission angle of 80
mr as a function of the detected proton momentum. Experimental
data are from \cite{azh5}. The calculations were made with the
deuteron wave functions for the Bonn B \cite{bonn} (dashed curve)
and the Paris \cite{paris} (dash-dotted curve) potentials. The
solid curve was calculated with the Karmanov's relativistic
deuteron wave function \cite{karm1}. }
\end{wrapfigure}

     In all previous papers concerned with the analysis of the
polarization observables of the $A(d,p)X$ reaction the deuteron
wave function has been presumed to be the superposition of the
$S$- and $D$-waves, each represented in the momentum space as a
product of angular and radial functions. In particular, this is
true for one of relativistic versions of theory --- relativistic
quantum mechanics \cite{uzikov}. This superposition  implies a
definite relationship between the transverse and longitudinal
components of the momentum of the internal motion of nucleons in a
deuteron \cite{azhyud1}.
     However, the dependence of the wave function on the transverse and
longitudinal components in the light-front dynamics may be
different from that dictated by the $S$- and $D$-wave combination.
The attention to this possibility was called in
\cite{schmidt,wong}, where the relativistic hard collision model
of composite hadrons \cite{sivers} was generalized to the case of
relativistic nucleus-nucleus collisions.
It is precisely this possibility that is explored in the present paper.

  The light-front dynamics \cite{dirac} has many important benefits
for the description of high-energy experiments. Probably the main
physical achievement of this approach is the prediction and
explanation \cite{miller1} of behaviour of the ratio of the
proton's elastic electromagnetic form-factors \cite{jones}. The
light-front dynamics is usually used also for description of the
deep inelastic phenomena. A special feature of this dynamics is
that the contribution of diagrams going back in time vanish.

     Difficulties of the light-front dynamics are in breaking of
rotational invariance as a result of selecting a particular
direction in space for the orientation of the light front. One
point related to this is that angular momentum operators $J_x,\
J_y$ in the light-front dynamics  become dynamical operators, i.e.
they are  dependent on an interaction. This leads to the
difficulties associated with the determination of the spin of a
composite system. References to papers devoted different aspects
of light-front dynamics can be found in the review \cite{miller2}.

     These difficulties with rotational invariance were
circumvented by Karmanov and coworkers \cite{karm1}.
They found the relativistic deuteron wave function with correct
internal spin.   This function
depends on two vector variables: on the momentum ${\bf k}$ of
nucleons in deuteron in their rest frame  and on the extra
variable ${\bf n}$ --- the unit normal to the light front surface.

\begin{wrapfigure}{l}{9cm}
\mbox{\epsfig{figure=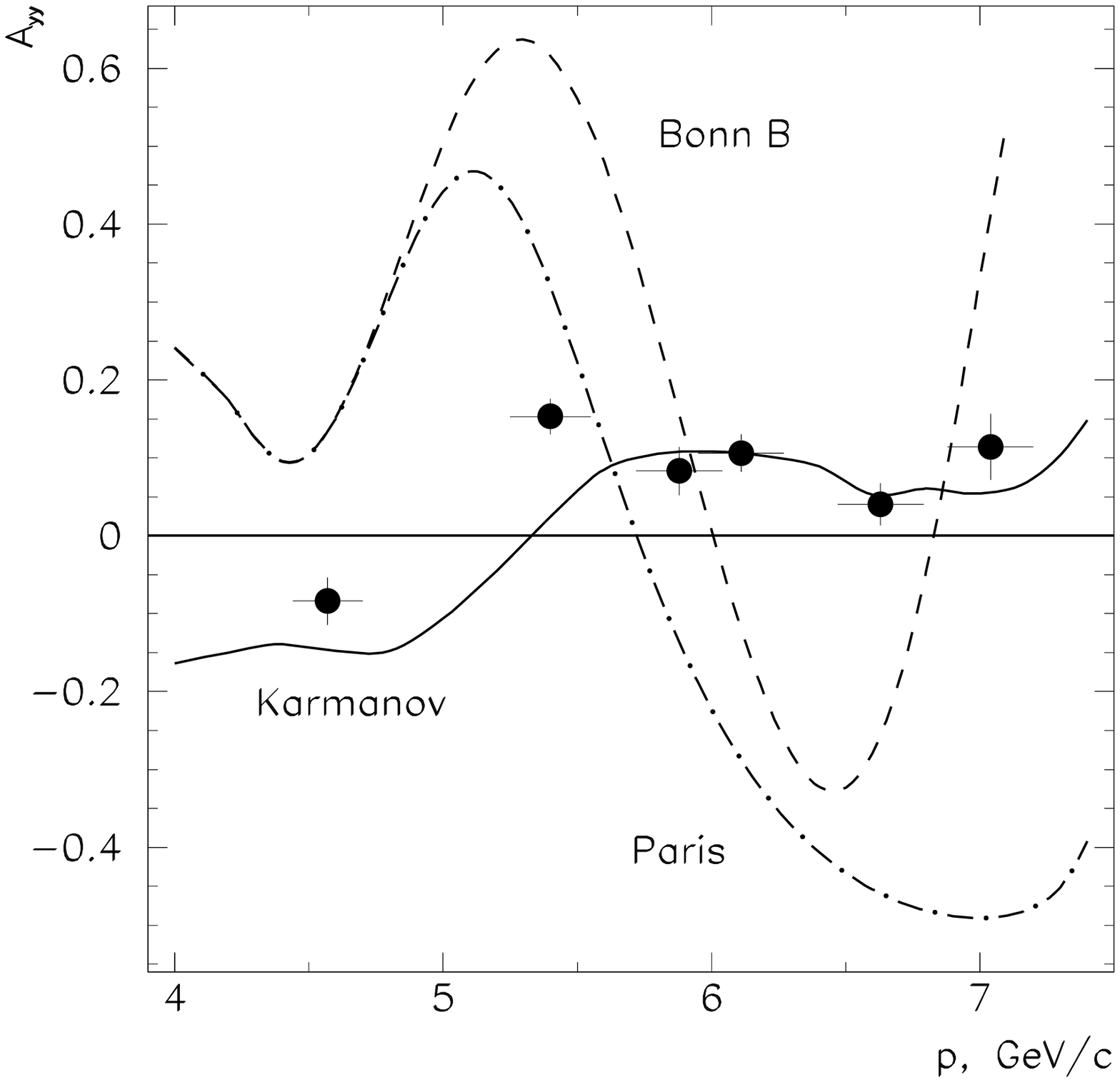,width=9cm,height=8cm}} 
{\small {\bf Fig. 2.} 
Parameter $A_{yy}$ of the reaction $^{12}C(d,p)X$ at 9 GeV/$c$ and
a proton emission angle of 85 mr versus  the detected proton
momentum. Experimental data from \cite{large_t}. The calculations
were made with the Bonn B \cite{bonn} (dashed curve), Paris
\cite{paris} (dash-dotted curve) and Karmanov's (solid curve)
deuteron wave functions \cite{karm1}. }
\end{wrapfigure}

The general statement by Karmanov is: the final results when
particles are on mass shell do not depend on the choice of the
plane of quantization. But this is right generally speaking only
in that case when one makes fully accurate calculations. Really
however one makes only approximate calculation and therefore it
arises some dependence on the choice of the front light surface.
We  play on this dependence. Since the direction along deuteron
beam is most evidently favoured we direct z-axis along beam. As a
result the wave function of relativistic deuteron with right spin
becomes nontrivially dependent on longitudinal and transverse
components of internal momentum and provides new possibilities in
the description of experimental data. We should like to point out
that without the accurate amplitudes of the process it is
difficult to put the serious argument in the favour  directing
z-axis along the deuteron beam. Therefore besides the intuitive
arguments, only the argument of the agreement with experiment for
several calculated  spin characteristics can be put forward.

Karmanov's wave function is determined by six invariant functions
instead of two ones in the non-relativistic case, each of them
depending on two scalar variables $k$ and $z = cos ({\bf
\widehat{k n}})$  and has the following form:
\begin{equation}
\Psi^M_{\sigma_2 \sigma_1} = w^\star_{\sigma_2} \psi^M({\bf k},
{\bf n}) \sigma_y w_{\sigma_1},
 \label{psim}
\end{equation}
where $M = 0,\, \pm 1$ are the projections of spin ${\bf J = 1}$
on the quantization axis, and
\begin{eqnarray}
{\bf \psi}({\bf k,\, n}) & = & \frac{1}{\sqrt{2}}{\bf \sigma} f_1 +
\frac{1}{2}\left[\frac{3}{k^2}{\bf k}({\bf k}\cdot {\bf \sigma}) -
{\bf \sigma}\right] f_2 + \frac{1}{2}\left[3{\bf n}({\bf n}\cdot
{\bf \sigma}) -
{\bf \sigma}\right] f_3 +  
\frac{1}{2k}[3{\bf k}({\bf n}\cdot {\bf \sigma})
\nonumber \\
& + & 3{\bf n}({\bf k}\cdot {\bf \sigma}) - 2{\bf \sigma}({\bf k}\cdot
{\bf n})] f_4 + \sqrt{\frac{3}{2}} \frac{i}{k}[{\bf k}\times {\bf
n}] f_5 + \frac{\sqrt{3}}{2k}[[{\bf k}\times {\bf n}]\times {\bf
\sigma}] f_6. \label{karmwf}
\end{eqnarray}
Here ${\bf \sigma}$ are the Pauli matrices, $w_{\sigma_1
(\sigma_2)}$ are the spin functions of non-relativistic nucleons,
and $f_1,...,f_6$ are the invariant about rotations functions of
the kinematical variables, that define the deuteron state. Here
\begin{equation}
k = \sqrt{\frac{m_p^2 + {\bf p}_T^2}{4x(1-x)} - m_p^2},\quad ({\bf
n} \cdot {\bf k}) = (\frac{1}{2} - x) \cdot \sqrt{\frac{m_p^2 +
{\bf p}_T^2}{x(1-x)}},
\end{equation}
where $x$ is the fraction of the deuteron longitudinal momentum
taking away by the proton in the infinite momentum frame.

     The invariant amplitude for the reaction $^1H(d,p)X$ in the light-front
dynamics is as follows
\begin{equation}
{\cal M}_a = \frac{{\cal M}(d \rightarrow p_1 b)} {(1-x)(M^2_d -
M^2(k))}{\cal M}(bp \rightarrow p_2 p_3),
\end{equation}
where ${\cal M}(d \rightarrow p_1 b)$ and ${\cal M}(bp \rightarrow
p_2 p_3)$ are the amplitudes of the deuteron breakup into the
particles $p_1,\, b$ and of the reaction $bp \rightarrow p_2 p_3$,
respectively. The ratio
\begin{equation}
\psi (x, p_{1T}) = \frac{{\cal M}(d \rightarrow p_1 b)} {M^2_d -
M^2(k)}
\end{equation}
is nothing but the wave function in the channel  $(b,\, N)$; here
$p_{1T}$ is the  component of momentum $p_1$ transverse to the $z$
axis, and $M^2(k)$ is given by
\begin{equation}
M^2(k) = \frac{m^2 + p^2_{1T}}{x} + \frac{b^2 + p^2_{1T}}{1-x},
\end{equation}
where $b^2$ is a four-momentum squared of the off-shell particle
$b$.

     The analyzing power of $T_{\kappa q}$ is given by
\begin{equation}
T_{\kappa q} = \frac{\int d\tau\,Sp\{{\cal M} \cdot t_{\kappa q}
\cdot {\cal M}^\dagger\}} {\int d\tau\,Sp\{{\cal M} \cdot {\cal
M}^\dagger\}},
\end{equation}
where $d\tau $ is the phase volume element, and the operator
$t_{2q}$ is defined by
$$
<m\,|\,t_{\kappa q}\,|\,m^\prime> = (-1)^{l-m}
<1\,m\,1\,-m^\prime\,|\,\kappa\,q>,
$$
with the Clebsh-Gordan coefficients
$<1\,m\,1\,-m^\prime\,|\,\kappa\,q>$.

     The final expression for the analyzing power has the form
\begin{eqnarray}
T_{2q}(\frac{p_{10}d\sigma}{d{\bf p_1}})_{un}  =
\frac{1}{2(2\pi)^3} \{\frac{I(b,p)}{I(d,p)\,(1-x)^2}
\rho_0(2,q)\,\sigma(bp \rightarrow X)
\nonumber \\
 +  \int \frac{dt\,d{\bf p}_{2T}}{2y(1-y)} \frac{I(b,p)}{(1-y)\,I(d,p)}
\rho_0(2,q) \frac{p_{10}d\sigma}{d{\bf p}_1}(bp \rightarrow p_2
p_3)\, [1 + {\bf P}<{\bf \sigma}>]\},
\label{t2q}
\end{eqnarray}
where $I(b,p),\ I(d,p)$ are the invariant fluxes of the
appropriate particles, $<{\bf \sigma}>$ is the vector analyzing
power of the $NN$-scattering, $\sigma(bp \rightarrow X)$ is
the total cross section of the $NN$-scattering, ${\bf P}$ is the
polarization vector of the nucleon in the deuteron that is
characterized by indices $(\kappa, q)$:
\begin{equation}
{\bf P} = Sp\{\rho(\kappa,q)\}/\rho_0(\kappa,q).
\end{equation}
The first term in the curly brackets of (\ref{t2q}) corresponds to
the case when the spectator proton is detected, and the second
term corresponds to the detecting of the proton scattered on the
target. The differential cross section for an unpolarized beam
entering in (\ref{t2q}) is given by
\begin{eqnarray}
(\frac{p_{10}d\sigma}{d{\bf p_1}})_{un}  =  \frac{1}{2(2\pi)^3}
\{\frac{I(b,p)}{I(d,p)\,(1-x)^2} \rho_0\,\sigma(bp \rightarrow p_2
p_3)
\nonumber \\
 +  \int \frac{dt\,d{\bf p}_{2T}}{2y(1-y)} \frac{I(b,p)}{(1-y)\,I(d,p)}
\rho_0 \frac{p_{10}d\sigma}{d{\bf p}_1}(bp \rightarrow p_2 p_3)\},
\end{eqnarray}
where
\begin{equation}
\rho_0  =  3[f_1^2 + f_2^2 + f_3^2 + f_2f_3(3z^2-1) + 4f_4(f_2 +
f_3)z + f_4^2(z^2+3) + (f_5^2 + f_6^2)(1-z^2)].
\end{equation}
If one introduces the density matrix in the spin space of the
nucleon $b$ at a given deuteron polarization characterized by
indices $(\kappa,\, q)$
\begin{eqnarray}
\rho_{\mu \mu^{\prime}}(\kappa, q) & = & \sum_{\nu,M,M^\prime}
\psi_M(\nu, \mu)(-1)^{1-M^\prime} <1\,M\,1\,-M^\prime|\,\kappa\,q>
\psi^\star_{M^\prime}(\nu, \mu^\prime) \nonumber \\
&  =  & \rho(\kappa, q) = \frac{1}{2}\rho_0(\kappa, q)(1 + {\bf
P}\cdot {\bf \sigma}),
\label{rhomm}
\end{eqnarray}
then the density matrices $\rho_0(\kappa,\, q)$ may be computed
from the relations
\begin{equation}
\rho_0(\kappa, q) = Sp\{\rho_{\mu, \mu^\prime}(\kappa,\, q)\}.
\end{equation}

     The results of calculations of tensor analyzing power $A_{yy}$ of
the reaction $^9Be(d,p)X$ at the initial deuteron momentum of 4.5
GeV/$c$ and a proton emission angle of 80 mr are compared with the
experimental data in Fig. 1.

     It is seen that the experimental data are qualitatively
properly reproduced using the Karmanov's relativistic deuteron
wave function as opposed to the calculations with the standard
deuteron wave functions \cite{paris,bonn}; the last  curves change
sign at the proton momentum $\sim$ 3.2 GeV/$c$.

     In Fig. 2 the experimental data on parameter $A_{yy}$ of the
reaction $^{12}C(d,p)X$ at the initial deuteron momentum of 9
GeV/$c$ and a proton emission angle of 85 mr are compared with the
calculations using different deuteron wave functions. It is seen
that the momentum dependence calculated with the relativistic
deuteron wave function is very close to the experimental points,
whereas the curves calculated with the standard non-relativistic
deuteron wave functions are in sharp contradiction with the data.

     Since the relativistic deuteron wave function \cite{karm1}
has a rather complicated appearance, the question arises, what
terms of this function help to describe qualitatively the
experimental data on the tensor analyzing power of the nuclear
fragmentation of the relativistic deuterons with emission of
protons with large transverse momenta? To answer this question,
the calculations of the momentum dependence of the parameter
$A_{yy}$ of the reaction $^{12}C(d,p)X$ at 9 GeV/$c$ and 85 mr
have been made, in which the terms $f_2,...,f_6$ of the function
(\ref{karmwf}) have been taken into account successively. The
results are shown in Fig. 3.

\begin{wrapfigure}{r}{9cm}
\mbox{\epsfig{figure=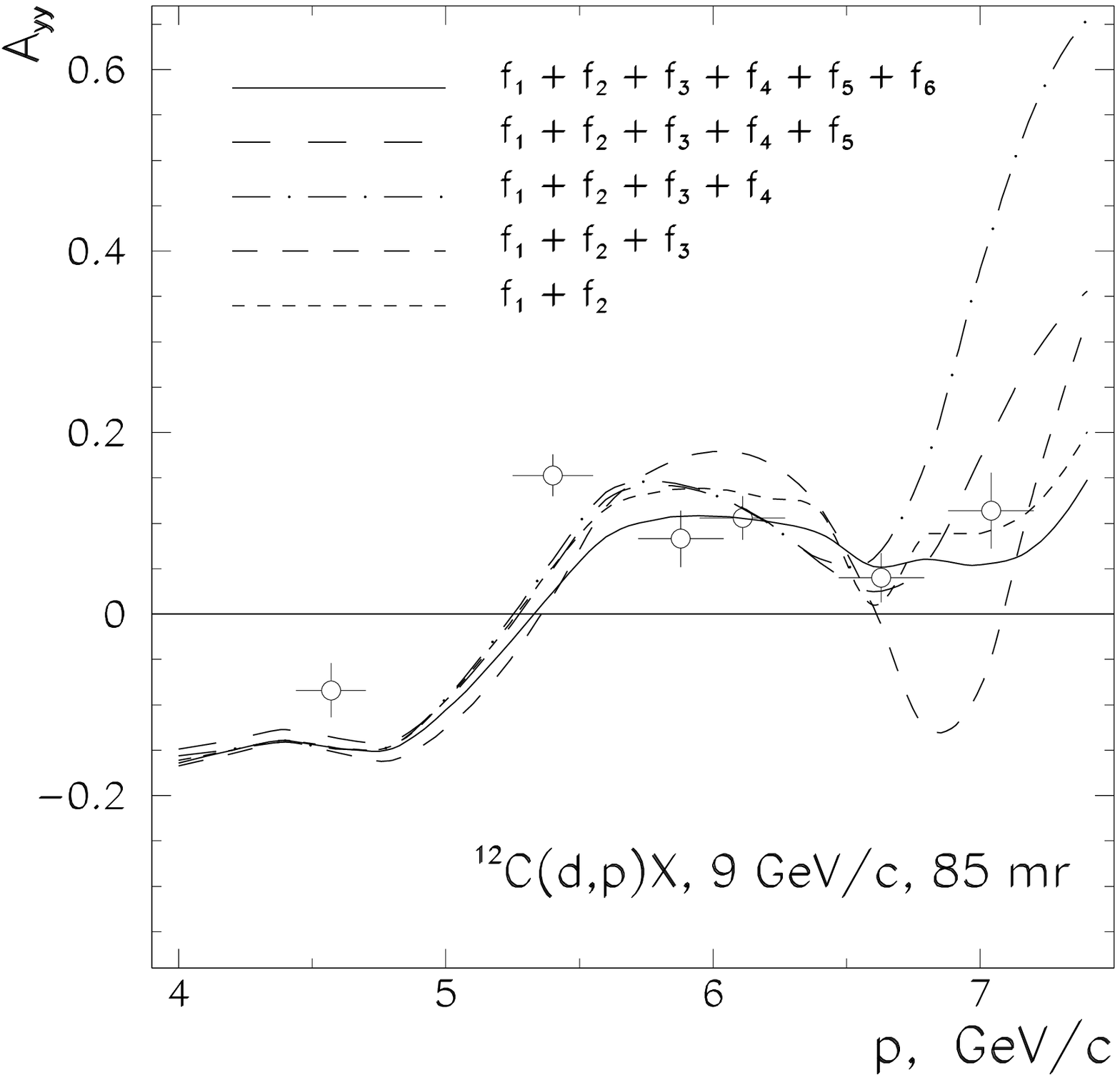,width=9cm,height=8cm}} 
{\small {\bf Fig. 3.}
Parameter $A_{yy}$ of the reaction $^{12}C(d,p)X$ at 9 GeV/$c$ and
a proton emission angle of 85 mr as a function of the detected
proton momentum. The different curves correspond to the successive
taking into account the terms of $f_i$ of the relativistic wave
function \cite{karm1}. }
\end{wrapfigure}

     It is seen that the two first terms of (\ref{karmwf}) give
the dominating contribution to the $A_{yy}$-dependence, and the
remaining terms give only corrections; the role of these
corrections increases with the momentum.

It is shown in
\cite{karm1} that in the non-relativistic limit the functions
$f_1$ and $f_2$ correspond to the $S$- and $D$-states of the
deuteron. Hence it follows that the relation between the $k_L$ and
${\bf k}_T$ in a moving deuteron differs essentially from that in
the non-relativistic case. The method of relativization proposed
by Karmanov et al. \cite{karm1} appear to reflect correctly this
relation.

     Two main conclusions may be made from this investigation.
At first, it turns out rather unexpectedly that up to small
relative distances corresponding to the internal momenta of
nucleons $k \sim 0.5 - 0.8$ GeV/$c$ the deuteron can be considered
as a two-nucleon system in the light-form of the quantum
mechanics. A similar conclusion was made in \cite{azh1} in
connection with the measurements of the momentum spectra of
protons emitted as a result of fragmentation of 9
GeV/$c$-deuterons in the region of proton transverse momenta of
0.5 - 1 GeV/$c$. Secondly, in the fragmentation process the
relativistic effects become significant very rapidly, and these
effects can be taken into account the most simple way through the
use of the light-front dynamics.

     This work was supported in part by the Russian Foundation for
Fundamental Research (grant No. 03-02-16224).

\end{document}